\documentclass[11pt]{article}

\usepackage[utf8]{inputenc}
\usepackage[T1]{fontenc}
\usepackage{mathptmx} 
\usepackage[a4paper,margin=0.9in]{geometry}
\usepackage[round,authoryear]{natbib}
\usepackage{graphicx}
\usepackage{amsmath}
\usepackage{microtype}
\usepackage[dvipsnames]{xcolor}
\usepackage{booktabs}
\usepackage{tabularx}
\usepackage{array}
\usepackage{authblk}
\usepackage[hidelinks]{hyperref}
\usepackage{url}

\title{Sensitivity of diffusion MRI to white matter pathology: a comparison between ROI-based effect size and TBSS analysis under different diffusion protocols, field strengths and post-processing strategies}

\author[1,2,3,*]{Evgenios N. Kornaropoulos}
\author[1]{Theodor Rumetshofer}
\author[9]{Anna Gard}
\author[4,7,8]{Linda Knutsson}
\author[9]{Niklas Marklund}
\author[1,5,6]{Pia C. Sundgren}
\author[1]{Markus Nilsson}

\affil[1]{Clinical Sciences, Diagnostic Radiology, Lund University, Lund, Sweden}
\affil[2]{GIGA CRC Human Imaging, University of Li\`ege, Li\`ege, Belgium}
\affil[3]{Aix Marseille Univ, CNRS, CRMBM, Marseille, France}
\affil[4]{Department of Medical Radiation Physics, Lund University, Lund, Sweden}
\affil[5]{Lund University BioImaging Center, Lund University, Lund, Sweden}
\affil[6]{Department of Medical Imaging and Physiology, Sk\aa ne University Hospital, Lund, Sweden}
\affil[7]{Russell H. Morgan Department of Radiology and Radiological Science, Johns Hopkins University School of Medicine, Baltimore, Maryland, USA}
\affil[8]{F.M. Kirby Research Center for Functional Brain Imaging, Kennedy Krieger Institute, Baltimore, Maryland, USA}
\affil[9]{Department of Clinical Sciences Lund, Neurosurgery, Lund University, Sk\aa ne University Hospital, Lund, Sweden}
\affil[*]{Corresponding author: ekornaropoulos@uliege.be}

\date{}

\begin{document}

\maketitle

\begin{abstract}
\noindent The purpose of this study was twofold: on one hand, to investigate the level to which more advanced diffusion MRI acquisitions and more complex processing increase the sensitivity to group-wise differences in white matter (WM) due to the presence of pathology, and on the other hand, to assess the impact of the choice of the statistical analysis of group-wise differences for evaluating sensitivity. Data were acquired with protocols adapted for DTI and DKI at 3T and 7T, in patients with systemic lupus erythematosus (SLE), in athletes with persistent post-concussive symptoms (PPCS), and in healthy controls (HC). Seven processing pipelines that included different combinations of state-of-the-art processing steps (denoising, Gibbs removal, eddy current and motion correction, and Gaussian smoothing) were applied. Regarding the statistical analysis for evaluating the sensitivity of DTI and DKI to WM pathology, two methods were followed: a region-of-interest-based (ROI-based) effect size analysis (72 WM tracts segmented by TractSeg), and a tract-based spatial statistics (TBSS) analysis, both aimed at detecting and evaluating group-level differences. In the ROI-based analysis, the sensitivity was quantified in terms of Cohen's $d$ for group comparison, while in the case of TBSS the sensitivity metric was the fraction of significant voxels in the WM skeleton. Results showed that for the PPCS case the two sensitivity analyses yielded different sensitivity results. While the ROI-based analysis did yield significant group-wise differences for both the ``SLE versus HC'' and ``PPCS versus HC'' cases, TBSS analysis did not reveal significant differences between PPCS and HC, but only between SLE and HC. Interestingly, both sensitivity analyses agreed that the diffusion protocol had a greater impact on overall sensitivity than that of the field strength and the post-processing strategy. Regardless of the applied post-processing pipeline, a fractional anisotropy map derived by DTI demonstrated a 0.26 increase in effect size and 25\% more significantly different voxels between SLE and HC than one derived by DKI, and a 0.2 increase in effect size along with 21\% more significantly different voxels at 3T than at 7T. Smoothing reduced sensitivity according to both the TBSS and the ROI-based analysis (up to 12\% according to TBSS in the SLE versus HC case). Overall, TBSS appears to be less suited for analyzing localized pathology, such as neurotrauma, in contrast to ROI-based analysis, which has not shown any bias with respect to the nature of the pathology. On another note, advanced diffusion protocols like 3T-DKI and 7T-DTI may need to be adjusted before surpassing conventional 3T-DTI in sensitivity to white matter pathology in SLE.

\vspace{0.5em}
\end{abstract}

\section{Introduction}

\noindent Disruption of white matter (WM) microstructure can be probed by diffusion MRI (dMRI). Basic techniques such as diffusion tensor imaging (DTI) characterize WM microstructure by parameters such as the mean, axial and radial diffusivity (MD, AD and RD respectively) and the fractional anisotropy (FA). Diffusion kurtosis imaging (DKI) extends DTI and provides additional parameters such as the mean, axial and radial kurtosis (MK, AK and RK respectively) \citep{jensen2005diffusional}, which are related to heterogeneity within the voxel \citep{henriques2020correlation}. The diffusion parameters serve as potential imaging biomarkers for assessing the extent of WM damage induced by neurodegeneration \citep{andica2020mr}. The sensitivity of those parameters to pathology may vary, depending not only on the diffusion protocol chosen (e.g.\ DTI or DKI) \citep{beck2021white,huang2022persistent,taha2022dki}, but also on the processing pipeline employed prior to extracting the diffusion parameters (e.g.\ a computationally intensive pipeline incorporating numerous state-of-the-art processing steps or simplifying the processing for speed and energy efficiency) \citep{beaudet2020age,harrison2020imaging}. Additionally, the magnetic field strength at which the diffusion protocol is implemented (e.g.\ an ultra-high field scanner or a conventional 3T scanner) is expected to play a crucial role too \citep{tannous2020stress,polders2011signal}. An evaluation of all these three aspects -- diffusion protocol, field strength, and processing pipeline -- would facilitate a more informed selection of methods for clinical and research studies.

\noindent In a previous study, \citet{kornaropoulos2022sensitivity} assessed the sensitivity of dMRI to WM pathology associated with systemic lupus erythematosus (SLE) in relation to the diffusion acquisition protocol, the magnetic field strength and the choice of the processing pipeline, using a region-of-interest-based (ROI-based) effect size analysis. The ROIs were defined as the WM tracts segmented by TractSeg \citep{wasserthal2018tractseg}. Effect sizes were analyzed using Cohen's $d$ \citep{coe2002s}. ROI-based methods deal with data in the subject space and thus avoid non-linear registrations to a template space, but may lack the level of detail provided by voxel-wise whole-brain approaches, like voxel-based morphometry (VBM) \citep{ashburner2000voxel} or its dMRI-optimized counterpart, tract-based spatial statistics (TBSS) \citep{bach2014methodological}. The statistical power of ROI-based analysis versus TBSS has been analyzed in the past \citep{hunter2019comparing,ly2015comparing,seo2013diffusion}, however, to the best of our knowledge, no study has yet assessed the impact of the diffusion protocol, the magnetic field strength and the preprocessing pipeline on the effectiveness of the analysis methods.

\noindent Moreover, the effectiveness of the analysis method may also depend on the type of WM damage being studied. For instance, the ROI-based method has shown greater efficacy in delineating the full extent of tissue abnormalities caused by neurotrauma such as in mild Traumatic Brain Injury (mTBI), in comparison to TBSS \citep{hunter2019comparing}. Conversely, TBSS provides a comprehensive overview of neurodegeneration in conditions like Parkinson's disease \citep{bergamino2021systematic} and Alzheimer's disease \citep{bergamino2020assessing}. In summary, it has yet to be fully explored how factors such as protocol, pipeline, and pathology affect the sensitivity of dMRI methods for detecting WM damage.

\noindent In this comparative study, we investigated the impact on sensitivity of diffusion MRI protocols (DTI and DKI) across two distinct types of white matter (WM) pathology: systemic lupus erythematosus (SLE), indicative of widespread WM damage, and persistent post-concussive symptoms (PPCS), representative of more focal WM damage. We also considered the impact of the magnetic field strength (3T and 7T) as well as the impact of various processing steps (denoising, Gibbs-ringing removal, eddy-current and motion correction, and smoothing, in different combinations). Our analysis employed both TBSS and ROI-based effect size analysis, utilizing Cohen's $d$ as the measure of effect size for the ROI-based analysis (as outlined in \citealp{kornaropoulos2022sensitivity}), and the fraction of significant voxels in the WM skeleton as the metric for the TBSS sensitivity. Cohorts of healthy controls were compared against the SLE patients and the PPCS athletes. For SLE, we assessed the congruence of TBSS results generated for this study with those from a ROI-based approach applied to the same cohort reported in \citet{kornaropoulos2022sensitivity}. Similarly, for PPCS, we compared TBSS results generated for this study with ROI-based analysis results on the same cohort reported in \citet{gard2024widespread}. By examining these two distinct pathologies within the same framework, our goal is to elucidate whether the selected sensitivity analysis method introduces any bias in the results for the respective pathologies.

\section{Methods}

\begin{figure}[p]
\centering
\includegraphics[width=\textwidth,height=0.82\textheight,keepaspectratio]{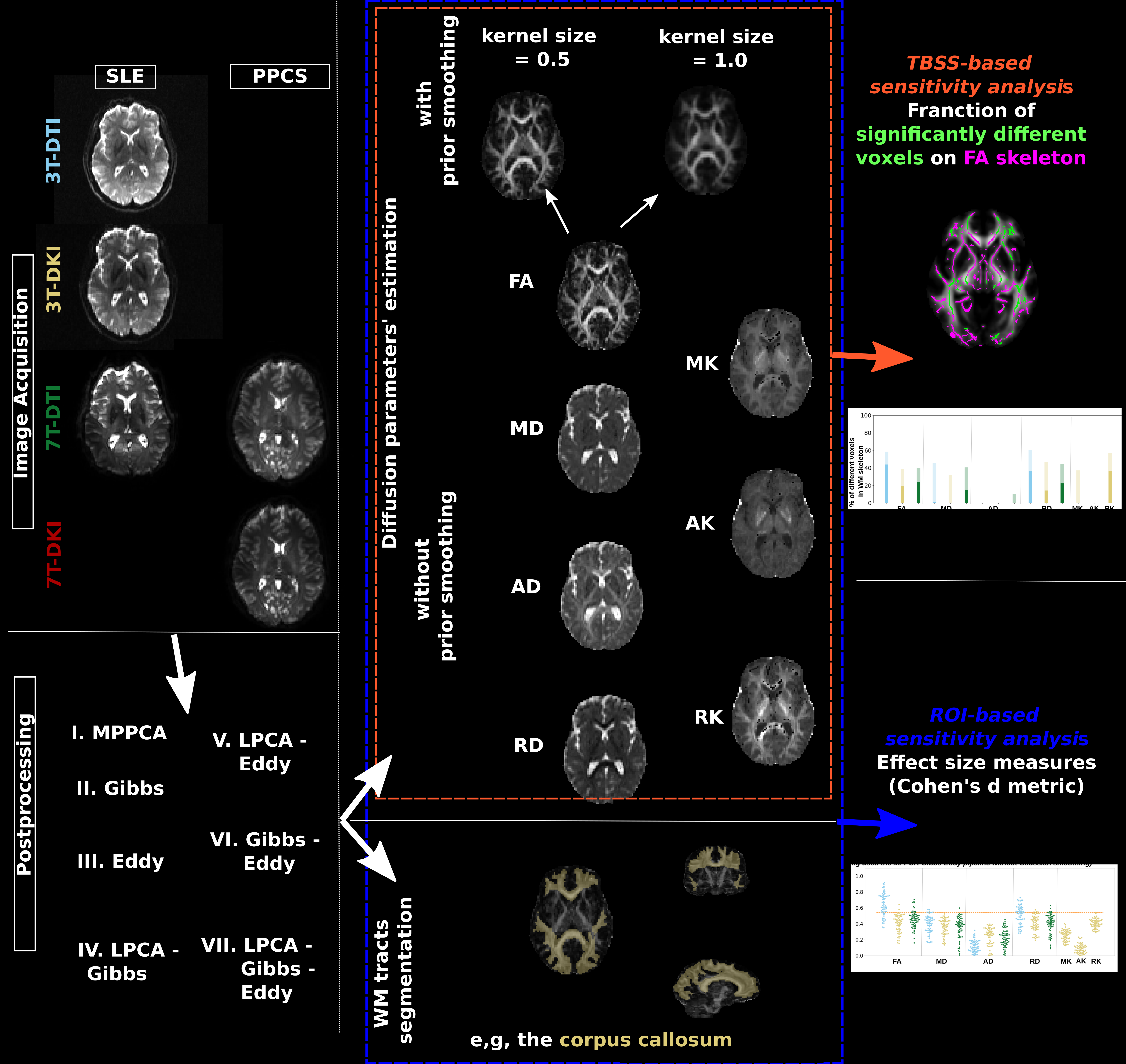}
\caption{The workflow. For the SLE study, the diffusion MRI (dMRI) data were acquired using three different acquisition protocols: 3T-diffusion tensor imaging (DTI), 3T-diffusion kurtosis imaging (DKI), and 7T-DTI. The three slices in the left-most column of diffusion-weighted MRI slice-images in the top-left block correspond to the same axial view of the dMRI volume with a b-value of 0 s/mm$^2$ in each of the three protocols. On the other hand, for the PPCS study the dMRI data were acquired using two different acquisition protocols: 7T-DTI and 7T-DKI. The two slices in the second-to-the-left column of images of the top-left block correspond to the same axial view of the dMRI volume with a b-value of 0 s/mm$^2$ in each of the two protocols. Subsequently, all the dMRI data from both studies were processed by seven different processing pipelines (bottom-left block). The processed dMRI data were then fitted to either a DTI or DKI model (depending on the initial acquisition) to extract dMRI parameters (middle-top block). To assess the impact of smoothing prior to model fitting, only in the case of the SLE study this fitting step was performed with different degrees of smoothing, ranging from no to substantial smoothing (middle-top images). Two different statistical analyses then followed: Regarding the ROI-based approach, the processed data were used by TractSeg to segment 72 major white matter (WM) tracts (middle-bottom block). The diffusion parametric maps combined with the segmented WM tracts were used as the input for the Cohen's $d$ analysis (blue panel). More specifically, we computed the mean and standard deviation of each of the diffusion parameters within each of the segmented WM tracts and those values were then used in the Cohen's $d$ formula to compute effect size values per diffusion parameter and per WM tract (see Equations 1 and 2). On the other hand, regarding the TBSS approach, the extracted FA maps from each subject were co-registered to the MNI template space (top-right block) and subsequently the rest of the parametric maps were transformed to that space in order to have voxel-wise correspondence across subjects and parametric maps. The creation of an FA skeleton followed. Finally the TFCE algorithm was applied in order to randomize the process of detecting the most significantly different points on the skeleton between healthy controls and either SLE patients or PPCS athletes.}\label{fig:1}
\end{figure}

The steps followed in this study are shown in Figure~\ref{fig:1}, and outlined in detail in this section.

\subsection{Participants}

\noindent Our analysis combined two studies: a study on SLE involving SLE patients and healthy controls (HC), and a study on PPCS involving athletes suffering from PPCS and HC. For the SLE study, a total of 106 female subjects were scanned. Out of these, 76 were SLE patients and 30 healthy controls. The Regional Ethical Review Board in Lund, Sweden approved the studies on 3T and 7T (\#2012/254, \#2014/778, \#2016/30, \#2019/01953) and all participants gave written informed consent prior to the examinations. None of the controls had a history of neurologic, neurodegenerative, or psychiatric disorder. The 106 subjects were examined with at least one of the acquisition protocols (see Table~\ref{tab:image_acquisition}). Not all subjects were scanned with all three protocols. However, the majority of subjects did overlap. For the PPCS study, a total of 41 male and female subjects were included in the analysis. Out of these, 20 were PPCS athletes and 21 healthy controls. 
The study was approved by the Regional Ethics Committee of Lund University, Lund, Sweden (Dnr 2017/1049) and conducted in accordance with the Declaration of Helsinki. All participants signed a written consent form. All of the PPCS athletes had previously been playing competitively at high national level. They also had a history of one sports-related concussion with ongoing PPCS for at least 6 months without symptom resolution at any time, with symptoms interfering with daily activities such as work, education, social life, and exercise. For more information on this cohort, we refer the readers to \citet{gard2024widespread}.

\subsection{Data acquisition}

\noindent Diffusion MRI data were acquired with four different protocols: a DTI and a DKI acquisition at a 3T Siemens Skyra scanner (3T-DTI and 3T-DKI) and a DTI and a DKI acquisition at a 7T Philips Achieva (7T-DTI and 7T-DKI). Not all diffusion protocols were applied to all subjects of the two studies due to time constraints. Due to the limitation in total scan time per patient, SLE patients were not scanned with the 7T-DKI protocol, while the protocol for the PPCS athletes did not include 3T-DTI and 3T-DKI. Scan time prioritizations also limited the encoding directions for 7T-DTI and 7T-DKI, which featured fewer encoding directions than the 3T-DTI and 3T-DKI protocols. The main acquisition parameters were adjusted for each protocol and system independently (see Table~\ref{tab:image_acquisition}). The objective of the adjustment was to minimize artefacts such as signal bias due to the rectified noise floor \citep{gudbjartsson1995rician}. Because high b-values attenuate the signal in DKI, the 3T-DKI protocol was acquired at a lower resolution than the 3T-DTI one.

\subsection{Processing pipeline}

\noindent Seven processing pipelines were built with the aim of analyzing the effect of post-processing on the sensitivity to WM pathology. Each of the pipelines included at least one of three main processing steps: denoising provided by a Marchenko--Pastur principal component analysis \citep{veraart2016denoising}, Gibbs-ringing artefact removal \citep{kellner2016gibbs}, and subject motion and eddy current-induced artefact correction from FSL \citep{jenkinson2012fsl}. Overall, the seven pipelines that were examined in this study were the following: \textit{MPPCA} for denoising, \textit{Gibbs} for removal of Gibbs-ringing artefacts, \textit{Eddy} for correction of motion and eddy current-induced distortions, \textit{MPPCA and Gibbs} in combination, \textit{MPPCA and Eddy} in combination, \textit{Gibbs and Eddy} in combination and \textit{MPPCA and Gibbs and Eddy} in combination. For reference, we also examined the effect on the sensitivity of not applying any post-processing at all (i.e.\ \textit{none}).

\noindent In addition, the impact of the level of Gaussian smoothing prior to the estimation of diffusion scalar metrics was examined, as this has been applied in many previous DTI and DKI studies \citep{delgado2017diffusion,maximov2019towards,nilsson2020tensor}. Here, the diffusion-weighted imaging data were smoothed using kernels with standard deviations ranging from 0 to 1 in units of 0.1. This part of our investigation used the data from the SLE study, as it had more subjects than the PPCS study.

\subsection{Segmentation of white matter tracts}

\noindent To obtain tract-specific parameter values, WM tract segmentation was performed using TractSeg \citep{wasserthal2019combined,wasserthal2018tractseg}. This is a convolutional neural network-based segmentation approach that automatically segments 72 major WM tracts in the native space of the diffusion-weighted images. The algorithm was pretrained on reference segmentations of tracts for 105 subjects from the Human Connectome Project \citep{sotiropoulos2013advances}. The main benefit of TractSeg is that it is both fast and accurate \citep{wasserthal2018tractseg}. A list with all 72 fiber tracts that it segments can be found online (\url{https://github.com/MIC-DKFZ/TractSeg}). Note that we computed mean parameters (e.g.\ mean FA) by using the tract probability-mask, thresholded at a value of 0.5.

\subsection{Diffusion parameter estimation}

\noindent We estimated the diffusion parameters using two approaches: DTI \citep{basser1994mr}, which provided FA, MD, AD, and RD, and DKI \citep{jensen2005diffusional} which in addition to the DTI metrics provided MK, RK, and AK. For DTI, the parameters were estimated through DTIFIT in FSL using weighted linear least squares. For DKI, we used the package dipy and its module \textit{DiffusionKurtosisModel}, again using weighted linear least squares. Example maps derived by the two methods (DTI and DKI) are shown in Figure~\ref{fig:1}.

\subsection{Effect size estimates}

\noindent We evaluated for each tract and diffusion parameter the groupwise difference between the SLE patients and the well-matched HC and between the PPCS athletes and the well-matched HC by Cohen's $d$ \citep{coe2002s}, which provides an effect size measure that can be easily compared across studies and populations \citep{dickie2015permutation,karayumak2019retrospective}, defined as:

\begin{equation}
    d = (u_{HC} - u_{PATH}) / s
\end{equation}

with
\begin{equation}
    s = \sqrt{((n_{HC} - 1){s_{HC}^2} + (n_{PATH} - 1){s_{PATH}^2}) / (n_{HC} + n_{PATH} - 2)}
\end{equation}
\noindent where $u_{HC}$ and $u_{PATH}$ are the mean values of the parameters within the tract for the HC and the group with the WM pathology (the SLE or the PPCS groups) respectively, and $s$ is the pooled standard deviation of the two groups. Moreover, $n_{HC}$ and $n_{PATH}$ are the sizes and $s_{HC}^2$ and $s_{PATH}^2$ the variances of the HC and the patient groups, respectively. To facilitate comparisons of effect sizes across parameters, we changed the sign of the effect size estimate in MD, RD, and AD so that all parameters had positive effect size estimates.

\noindent To assess statistical significance, we note that the 95\% confidence interval for Cohen's $d$ -- in the absence of a true effect (an effect size of zero) for group sizes of 56 and 20, as in 3T-DKI -- spans the range where the magnitude of $d$ is below 0.54. Effect sizes larger than this can thus be considered significant on a 5\% significance level \citep{hedges2014statistical}. In the case of both 3T-DTI and 7T-DTI that level of significance was set to 0.53, based on the corresponding sizes of the cohorts. Effect sizes were estimated in 72 tracts and so a multiple comparisons problem arises. We did not correct for this, but note that a Bonferroni-like correction can be applied \citep{holm1979simple,armstrong2014use,abdi2007bonferroni}. Correcting for 72 independent tests corresponds to using the 99.9\% confidence interval ($1 - 0.05/72$), which in turn corresponds to a minimum threshold for a significant effect at $d = 0.84$. Another way to approach the multiple comparisons problem is to note that in the absence of a true effect in all 72 tracts, the probability to still identify more than seven tracts as significant on the uncorrected significance level (5\%) is less than 5\% \citep{mac2011detection}. This means that observing more than seven tracts as significant on the uncorrected level indicates a true effect in at least some of those tracts.

\subsection{Statistical analysis on sensitivity through TBSS}

\noindent TBSS analysis was performed using FSL \citep{bach2014methodological}. The FA was used for skeletonization. Tests were performed for the FA as well as on the MD, AD and RD, and for 3T-DKI only also on the mean, axial and radial kurtosis (MK, AK and RK respectively). The target image used in the registrations was the pre-defined target provided by FSL (i.e.\ the FMRIB58\_FA standard-space image), using the \texttt{-T} flag option. Therefore, all images were co-aligned to a $1\times{1}\times{1}$\,mm$^3$ standard space. Statistical analysis was carried out by performing permutation tests (5000 times) \citep{winkler2014permutation}. Multiple comparison correction was done using threshold-free cluster enhancement (TFCE) \citep{smith2009threshold}. A family-wise error corrected $P$ value was ascribed to each cluster of voxels comprising the WM skeleton. The significance threshold was set to $P < 0.05$. Having computed the clusters of significantly different voxels upon the WM skeleton, the fraction of the sum of all the significant voxels to the sum of all the voxels belonging to the WM skeleton was used as the sensitivity metric, using the following formula:

\begin{equation}
    \mathrm{sensitivity} = 100 \cdot \frac{\sum_{i=0}^{M}N_i}{K},
\end{equation}

\noindent where $M$ is the total number of clusters found, each having $N_i$ significantly different voxels, and $K$ is the total number of voxels upon the WM skeleton.

\section{Results}

\subsection{Effects of pipelines on sensitivity to SLE}
\begin{figure}[p]
\centering
\includegraphics[width=\textwidth,height=0.82\textheight,keepaspectratio]{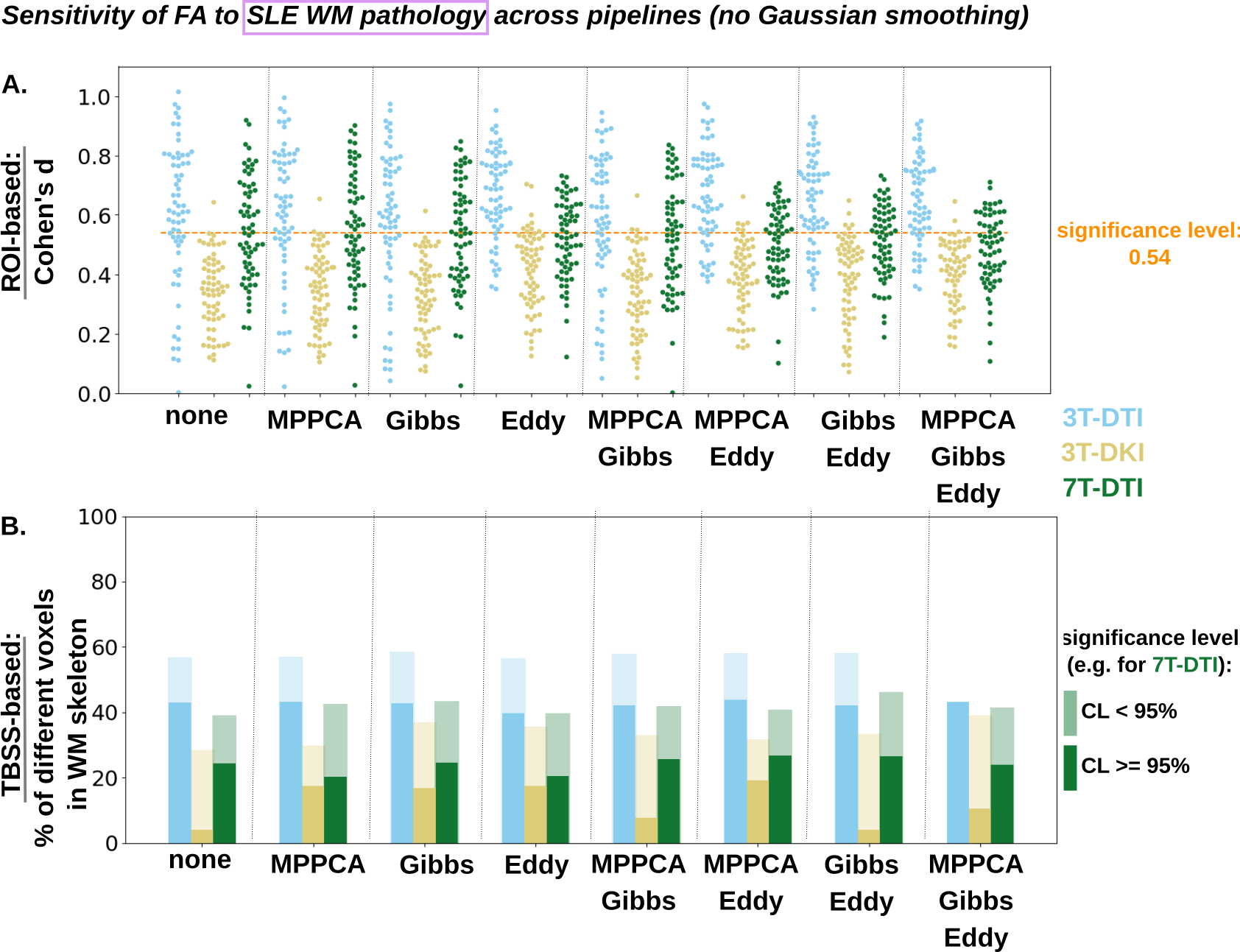}
\caption{Impact of the processing pipeline. The impact of the processing pipelines (x-axis) is evaluated through effect size estimates (Cohen's $d$ scores, y-axis of top plot) and percentage of different points on the skeleton between HC and SLE patients.}\label{fig:2}
\end{figure}

\begin{figure}[p]
\centering
\includegraphics[width=\textwidth,height=0.82\textheight,keepaspectratio]{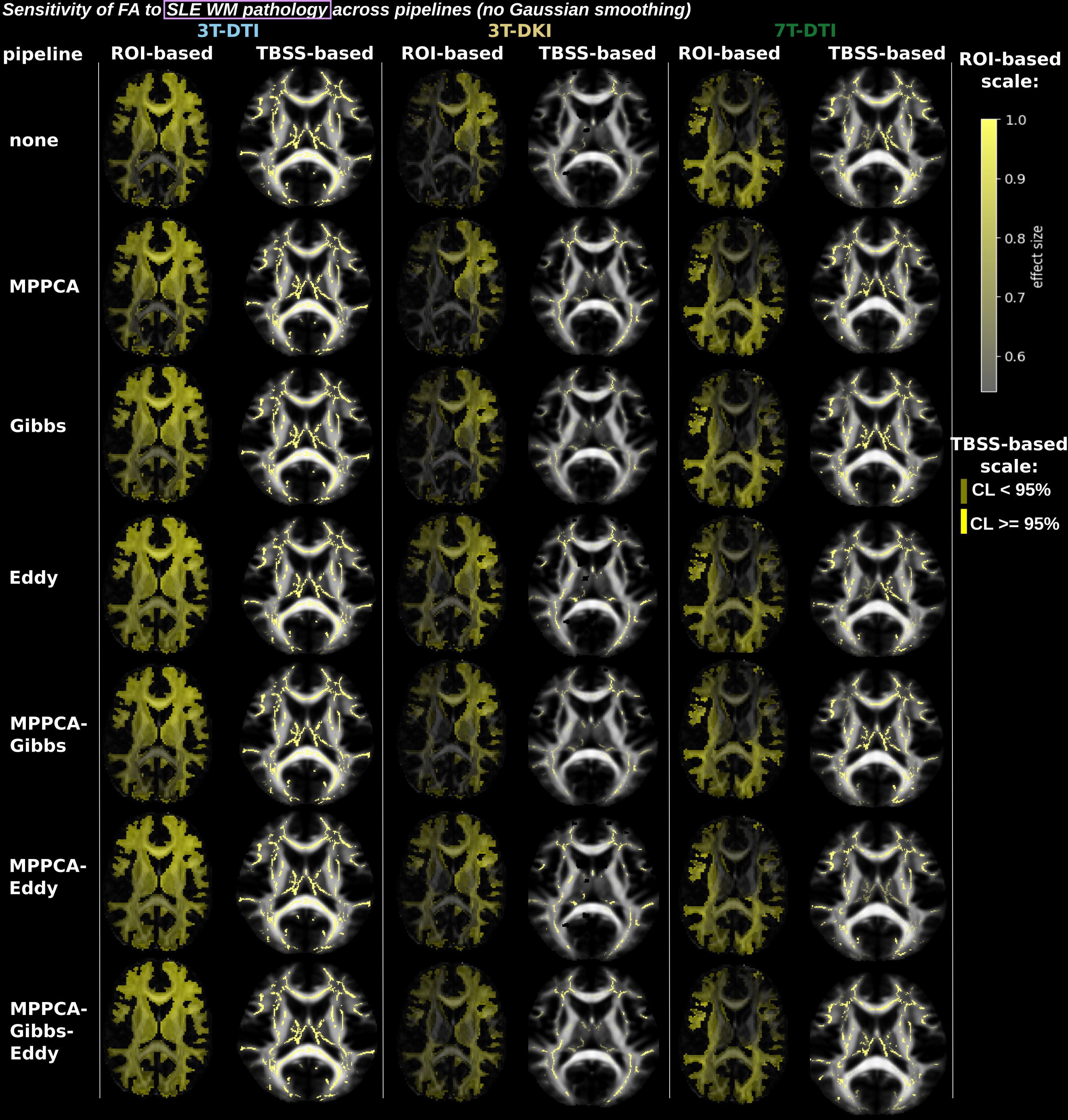}
\caption{Impact of the processing pipeline across brain regions. The impact of the processing pipelines is evaluated through effect size estimates (Cohen's $d$ scores for the ROI-based approach) and percentage of different points on the skeleton between HC and SLE patients (for the TBSS-based approach).}\label{fig:3}
\end{figure}

\noindent Both the ROI-based and the TBSS-based sensitivity approach demonstrated a minor effect of the choice of the processing pipeline on sensitivity for FA (Figure~\ref{fig:2}). When following the ROI-based approach, the aforementioned result was observed in all the diffusion parameters. However, in the case of TBSS, the incorporation of MPPCA, Gibbs, and especially Eddy in the case of FA and RK derived from the 3T-DKI data increased the number of significantly different voxels in TBSS. The fact that Eddy was beneficial for 3T-DKI-derived FA can be observed in Figure~\ref{fig:3} as well, which shows that the TBSS analysis of the 3T-DKI data processed by Eddy demonstrated many more significantly different voxels (yellow color) than the analysis of 3T-DKI data without any processing at all. In fact, Eddy substantially increased the sensitivity to SLE WM pathology in the case of 3T-DKI-derived RD too. In contrast to the clearly positive effect of Eddy, the incorporation of MPPCA or Gibbs affected sensitivity to a lesser degree in any of the other acquisition protocols. Overall, FA derived by 3T-DTI was the parameter with the highest sensitivity, regardless of the post-processing strategy followed.

\subsection{Effects of Gaussian smoothing on sensitivity to SLE}
\begin{figure}[p]
\centering
\includegraphics[width=\textwidth,height=0.82\textheight,keepaspectratio]{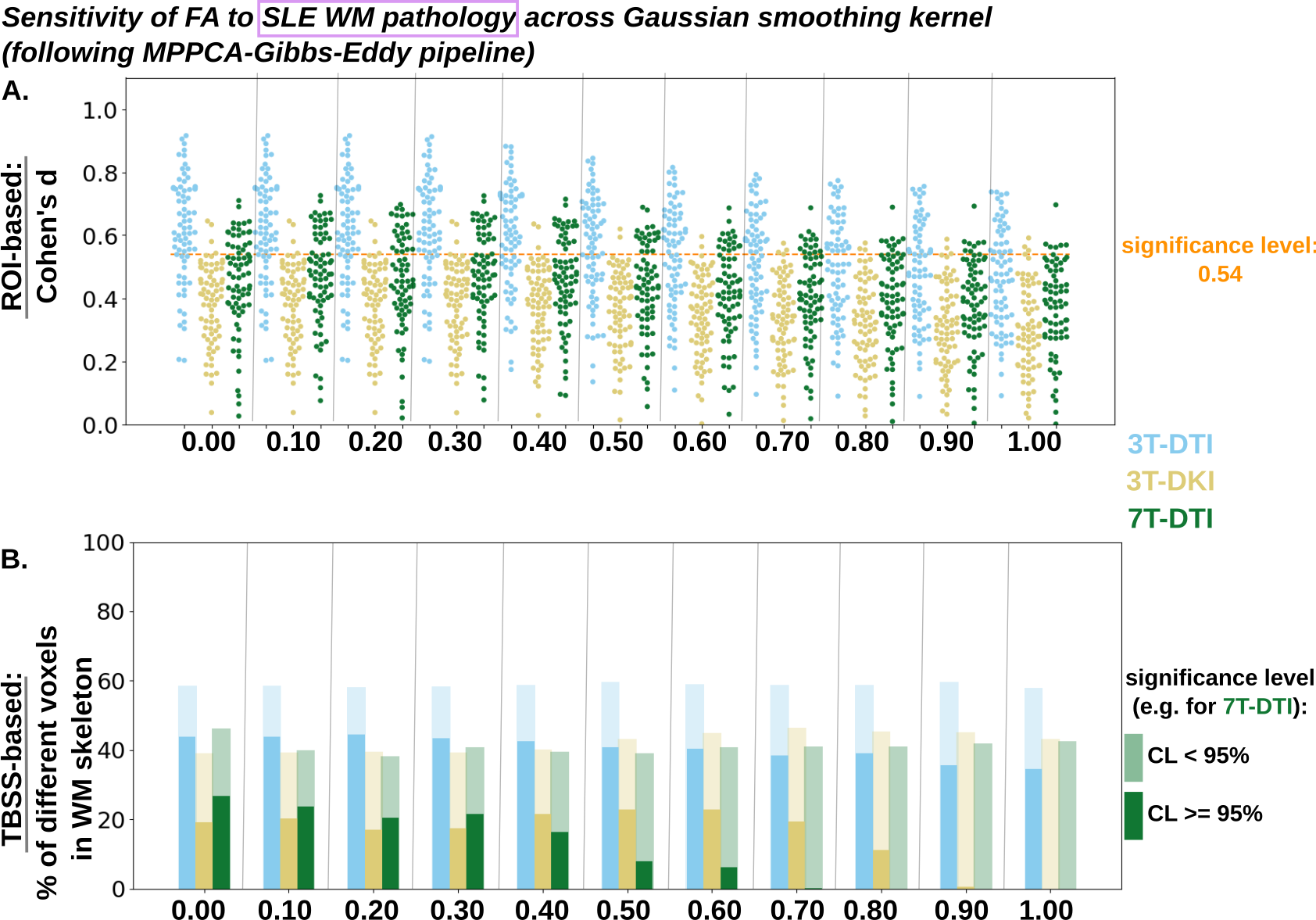}
\caption{Impact of Gaussian smoothing. The impact of the size of the Gaussian smoothing kernel (x-axis) is evaluated through effect size estimates (Cohen's $d$ scores, y-axis of top plot) and percentage of different points on the skeleton between HC and SLE patients.}\label{fig:4}
\end{figure}

\begin{figure}[p]
\centering
\includegraphics[width=\textwidth,height=0.82\textheight,keepaspectratio]{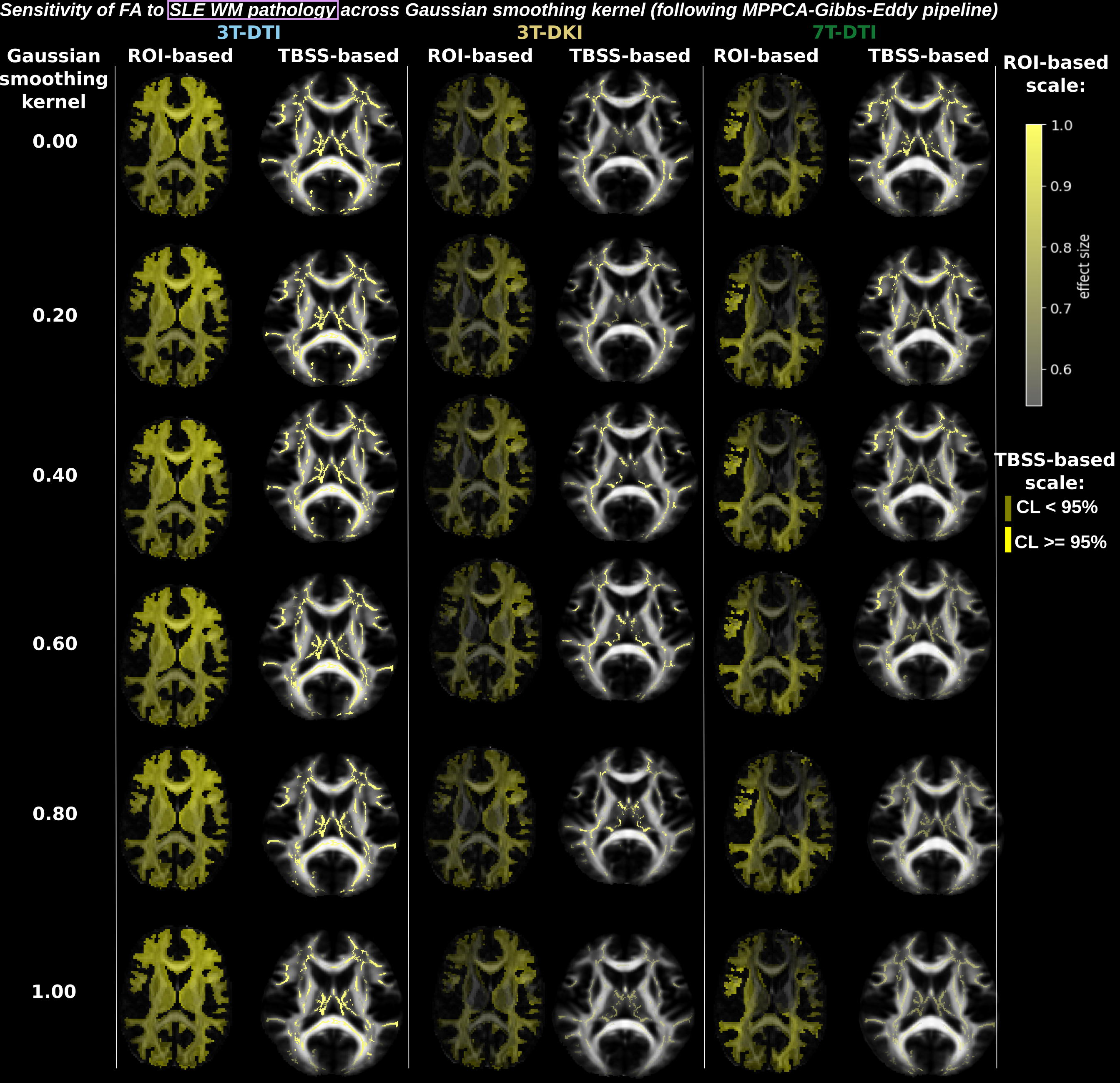}
\caption{Impact of Gaussian smoothing across brain regions. The impact of the size of the Gaussian smoothing kernel is evaluated through effect size estimates (Cohen's $d$ scores for the ROI-based approach) and percentage of different points on the skeleton between HC and SLE patients (for the TBSS-based approach).}\label{fig:5}
\end{figure}

\noindent Figures~\ref{fig:4} and \ref{fig:5} show sensitivity estimates across the different sizes of an applied Gaussian smoothing kernel, where increasing smoothing results in a reduction of significantly different voxels (magenta color) on the WM skeleton. Large-scale smoothing reduced the sensitivity, not only for FA and RK but for all diffusion parameters. Slight smoothing had a limited positive effect for some protocol-parameter combinations.

\subsection{Effects of magnetic field strength and acquisition protocol on sensitivity to SLE}
\begin{figure}[p]
\centering
\includegraphics[width=\textwidth,height=0.82\textheight,keepaspectratio]{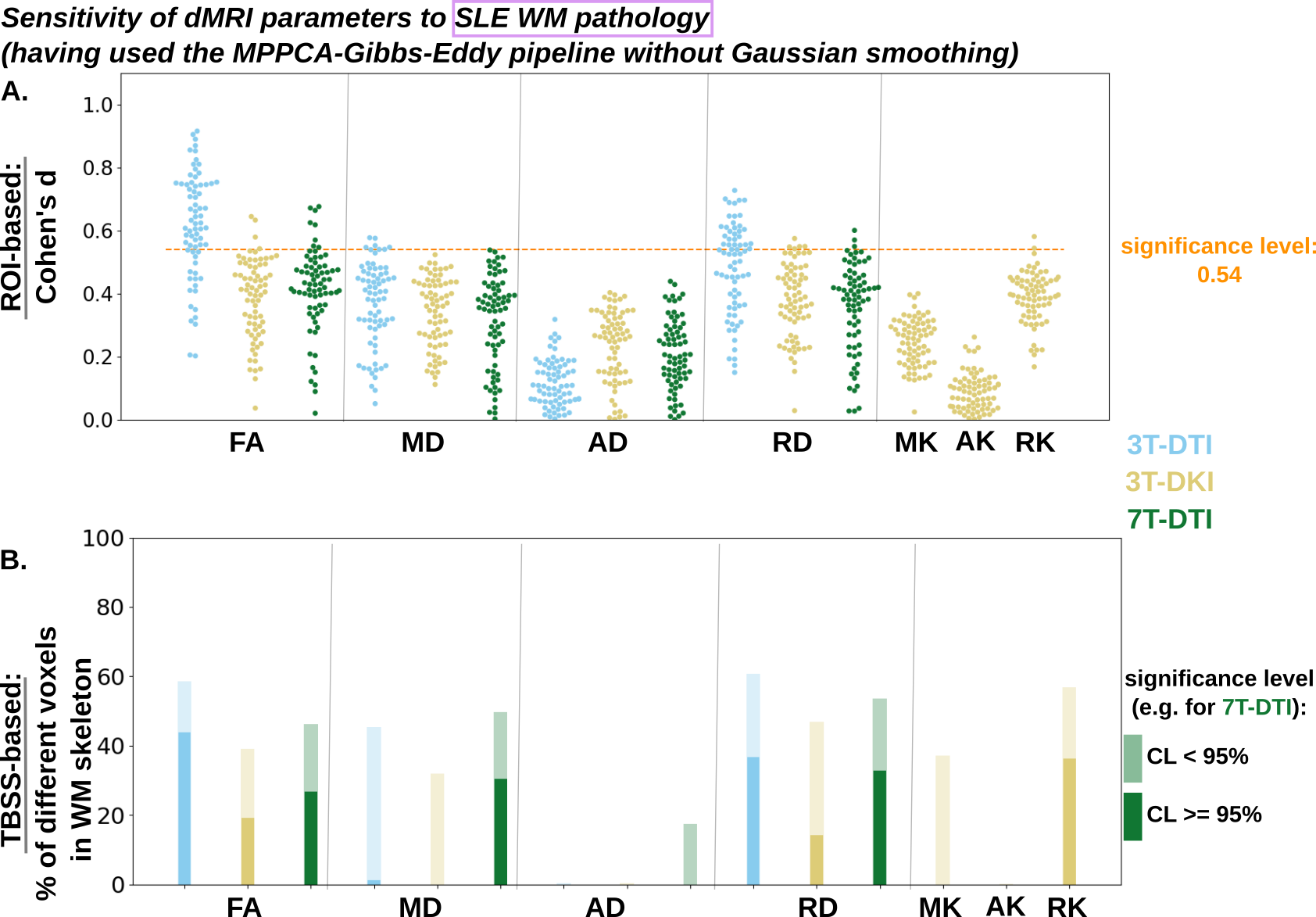}
\caption{Evaluation of the acquisition protocol on the effect size. Columns show different diffusion parameters obtained with different protocols 3T-DTI, 3T-DKI, or 7T-DTI, shown in cyan, gold, and green, respectively. Top row: The effect size based on the mean of each diffusion parameter was examined. Pipeline VII was used in all cases. No smoothing was applied. The swarm and box plots show the distribution and quartiles (together with the median) of Cohen's $d$ values among tracts, respectively. Bottom row: The percentage of different voxels on the FA skeleton between HC and SLE patients was evaluated across the different diffusion parameters derived by either 3T-DTI or 3T-DKI or 7T-DTI.}\label{fig:6}
\end{figure}

\begin{figure}[p]
\centering
\includegraphics[width=\textwidth,height=0.82\textheight,keepaspectratio]{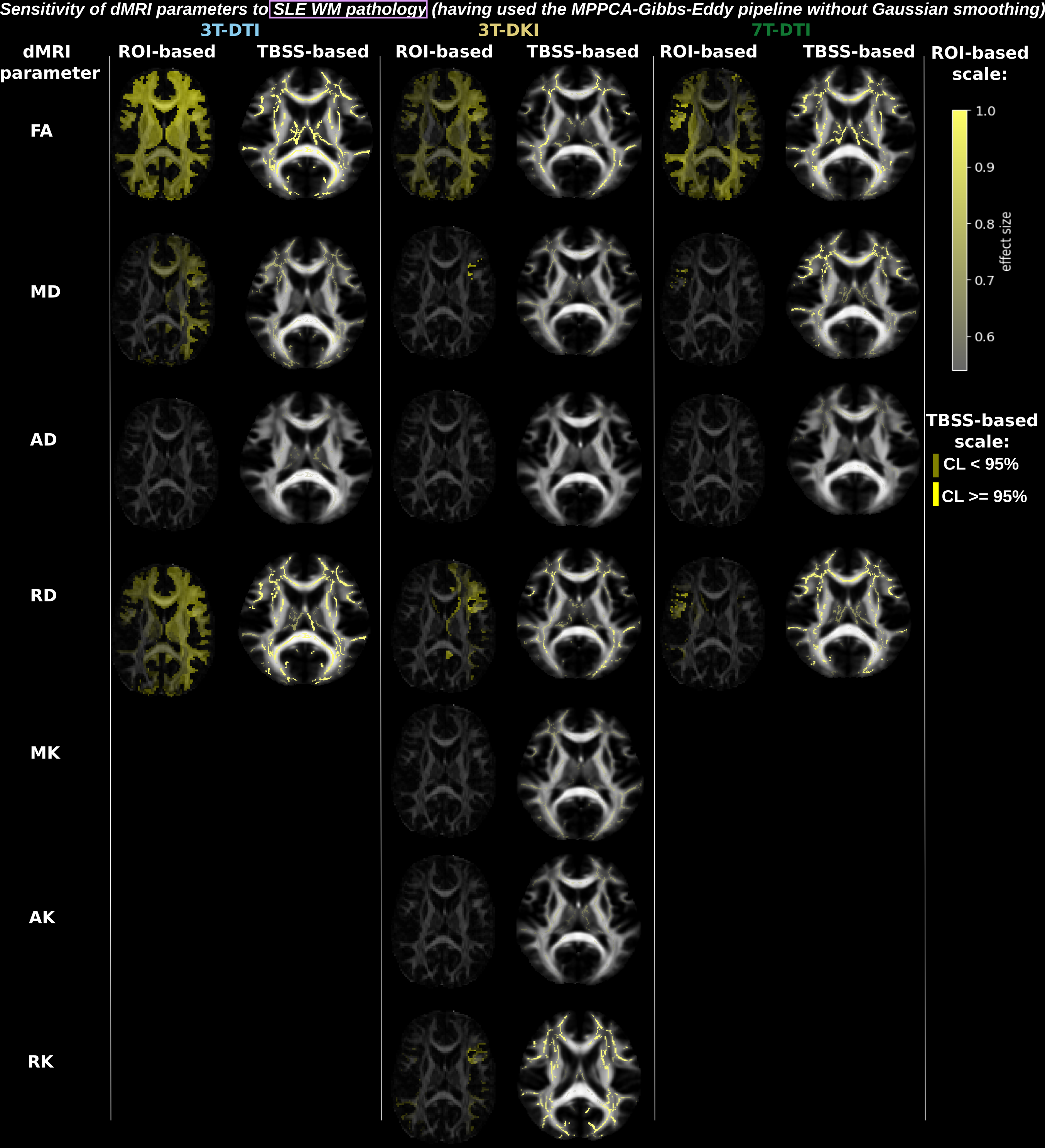}
\caption{Evaluation of the acquisition protocol on the effect size across brain regions in case of the SLE study.}\label{fig:7}
\end{figure}

\noindent Figures~\ref{fig:6} and \ref{fig:7} show the results of the sensitivity analysis to SLE pathology for all diffusion parameters derived by 3T-DTI, 3T-DKI and 7T-DTI using both the ROI-based and the TBSS approach. Regardless of the sensitivity analysis approach (ROI-based or TBSS), we found that three parameters demonstrated significant differences between healthy controls and SLE patients (FA, RD and RK). Controls had consistently higher values of FA and RK compared to SLE patients, while the opposite was observed for RD. In short, the TBSS and ROI-based results for SLE are similar, which corroborates the findings of \citet{kornaropoulos2022sensitivity}.

\subsection{Effects of magnetic field strength and acquisition protocol on sensitivity to PPCS}
\begin{figure}[p]
\centering
\includegraphics[width=\textwidth,height=0.82\textheight,keepaspectratio]{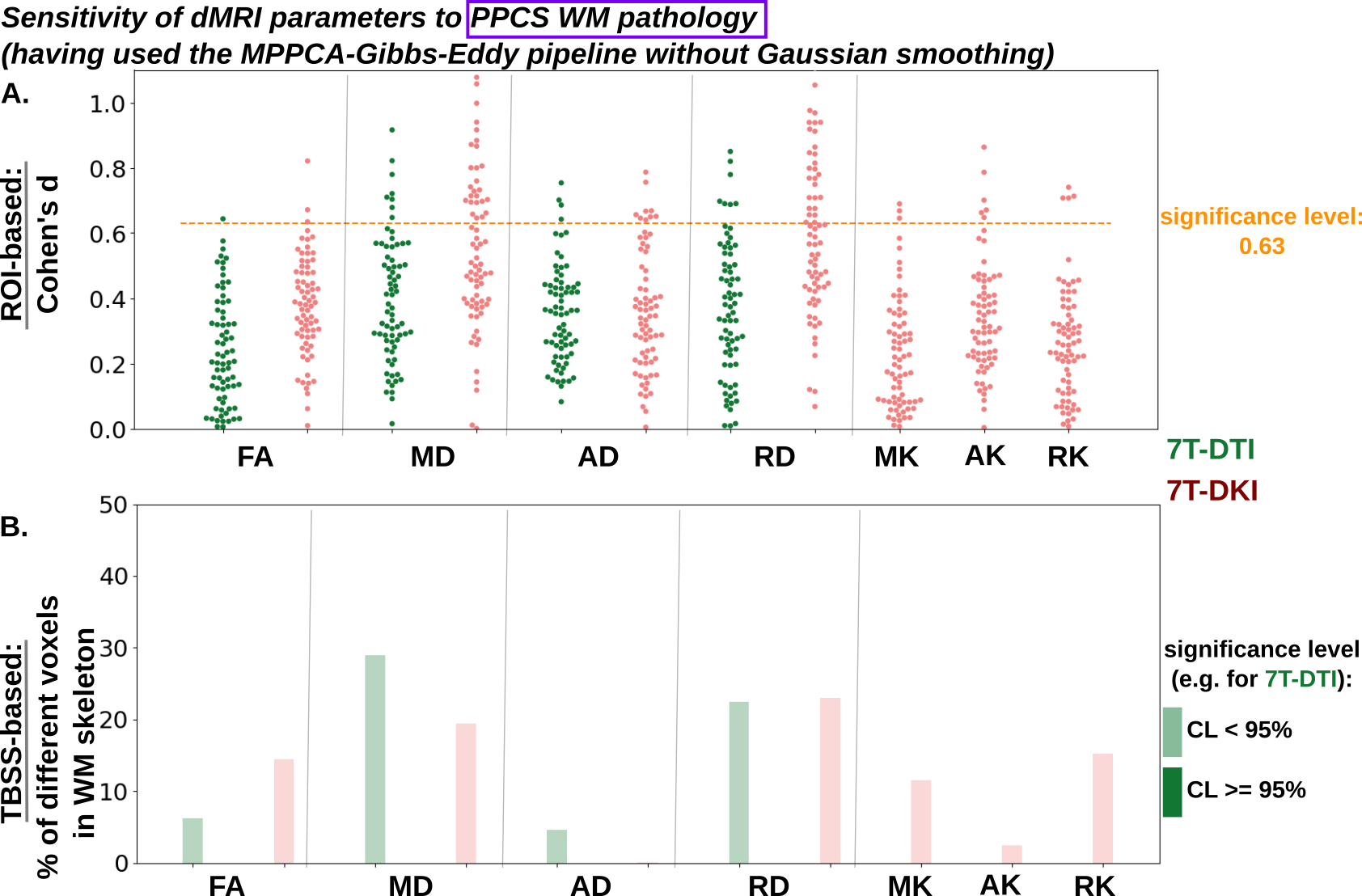}
\caption{Evaluation of the acquisition protocol on the effect size for the PPCS study. Columns show different diffusion parameters obtained with different protocols 7T-DTI or 7T-DKI, shown in green and lightcoral, respectively. Top row: The effect size based on the mean of each diffusion parameter was examined. Pipeline VII was used in all cases. No smoothing was applied. The swarm and box plots show the distribution and quartiles (together with the median) of Cohen's $d$ values among tracts, respectively. Bottom row: The percentage of different voxels on the FA skeleton between HC and PPCS athletes was evaluated across the different diffusion parameters derived by either 7T-DTI or 7T-DKI.}\label{fig:8}
\end{figure}

\begin{figure}[p]
\centering
\includegraphics[width=\textwidth,height=0.82\textheight,keepaspectratio]{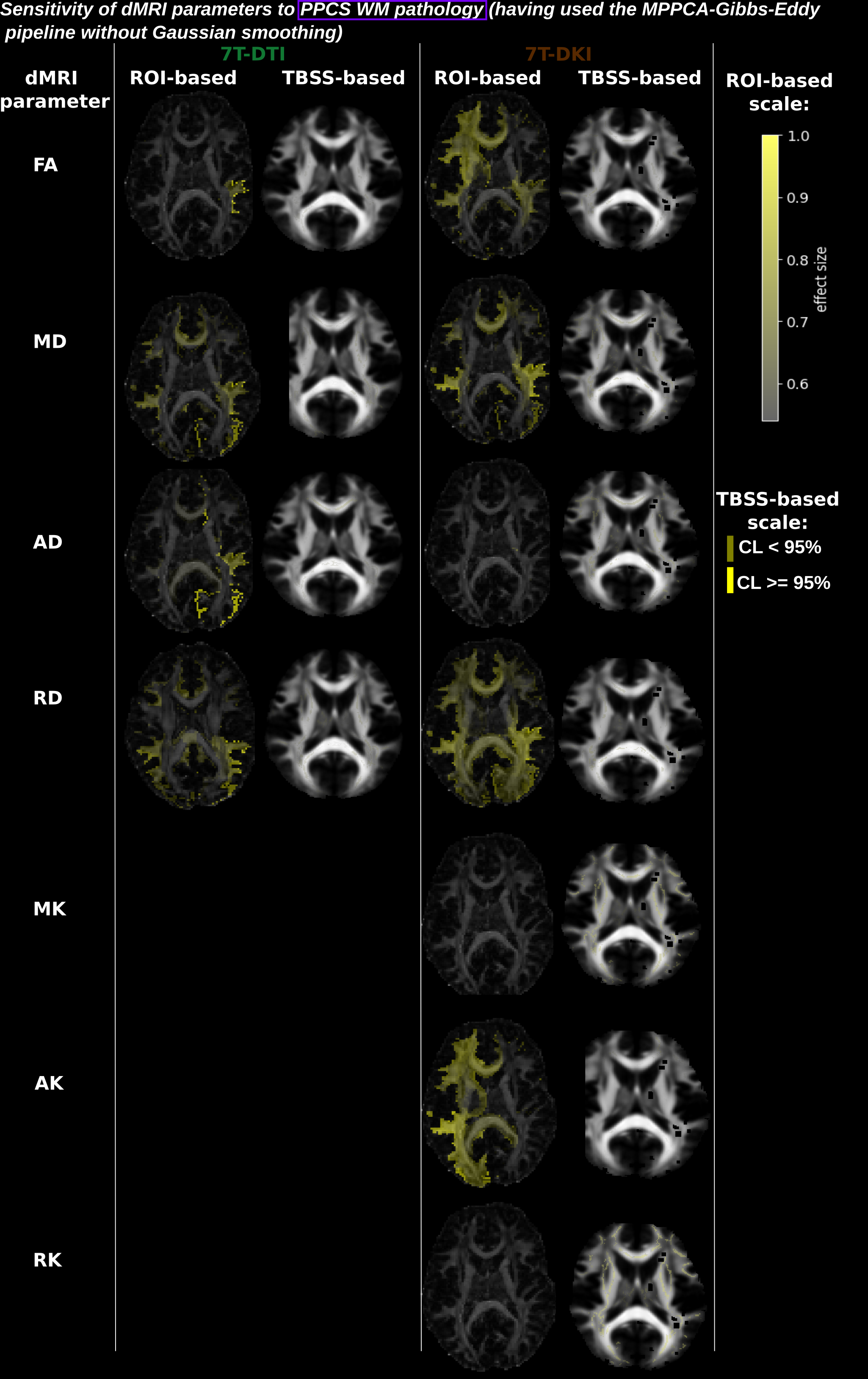}
\caption{Evaluation of the acquisition protocol on the effect size across brain regions in case of the PPCS study.}\label{fig:9}
\end{figure}

\noindent Figures~\ref{fig:8} and \ref{fig:9} show the results of the sensitivity analysis to PPCS pathology for all diffusion parameters derived by 7T-DTI and 7T-DKI using both the ROI-based and the TBSS approach. The MD and RD parameters demonstrated high sensitivity when following the ROI-based approach, especially for the case of 7T-DKI. However, that was not the case for TBSS, as no diffusion parameter ended up having any significantly different voxel on the WM skeleton between HC and PPCS athletes.

\section{Discussion}

\noindent The most important observation from our comparative analysis is that the choice of the dMRI sensitivity analysis approach itself can introduce systematic bias into the outcomes, depending on both the data characteristics and the underlying pathology. Our analysis combined two studies: one in SLE with a relatively large sample (76 patients, 30 controls) and one in PPCS with a smaller sample (20 athletes, 21 controls). Both the TBSS and the ROI-based approaches detected robust effects in SLE, but only the ROI-based method yielded significant findings in PPCS. The discrepancy raises a critical methodological concern: is TBSS particularly susceptible to false negatives when applied to small, low-powered datasets?

\noindent Most studies directly comparing TBSS with tract-based or ROI-based approaches report results broadly similar to our SLE findings when applied to sufficiently powered datasets \citep{grinberg2017diffusion,kamagata2014preliminary,xiong2019brain,yang2021white}. However, \citet{hu2023microstructure} reported findings comparable to our PPCS results: in a relatively small cohort of children with ADHD (27 patients, 22 controls), TBSS results did not survive correction for multiple comparisons, whereas Automated Fiber Quantification (AFQ) tract-based analysis \citep{yeatman2012tract} detected significant group differences. This has important implications for study design: the fact that TBSS failed to detect PPCS differences at $n \approx 20$ subjects per group, whereas ROI-based analysis succeeded, suggests that TBSS may require larger samples to achieve equivalent power, particularly when dealing with subtle or spatially heterogeneous pathology such as PPCS. This aligns with previous work comparing ROI-based and voxel-wise meta-analyses, where researchers found that ROI-based meta-analysis of DTI studies in mild and moderate TBI yielded effect sizes roughly twice those observed with voxel-wise analyses, consistent with the reduced statistical power of voxel-wise approaches after multiple-comparison correction \citep{hunter2019comparing}. Although several studies have used TBSS and ROI-based approaches in parallel (Table~\ref{tab:articles_tbss_and_roi}), they typically treat the methods as complementary rather than explicitly comparing their efficacy. Our results highlight the need for systematic evaluation of method-specific sensitivity, particularly in underpowered studies.

\noindent The discrepancy observed in the PPCS cohort also underscores structural limitations of TBSS. TBSS relies on projecting individual diffusion maps onto a common white-matter skeleton \citep{smith2006tract}, which inherently emphasizes deep, coherent tracts. Consequently, focal abnormalities at the gray--white matter boundary or in superficial white matter may be excluded from the skeleton and thus underdetected \citep{bach2014methodological}. This limitation is particularly relevant in traumatic brain injury, where diffuse axonal injury often produces spatially heterogeneous and focal pathological changes across white matter tracts \citep{armstrong2024white}. Nonlinear registration further compounds the issue by smoothing or displacing subtle lesions, thereby diminishing sensitivity in small samples. In contrast, ROI- or tract-based analyses preserve native-space tract information and may therefore better capture focal pathology, as seen in our PPCS dataset.

\noindent It is also possible that the divergent findings reflect the underlying pathology rather than cohort size. SLE is characterized by chronic, diffuse WM injury, often involving demyelination, axonal degeneration, and microvascular damage \citep{correa2018white,ercan2015multimodal}. Such widespread changes are more likely to align with TBSS's skeleton-based framework. Conversely, PPCS is typically linked to focal diffuse axonal injury (DAI), microhemorrhages, and localized gliosis \citep{haber2021imaging}, processes that produce spatially heterogeneous, small-scale abnormalities less amenable to TBSS detection. \citet{karlsen2019diffusion} partially support this interpretation: in acute mild TBI (22 patients, 22 controls), both TBSS and ROI-based DKI metrics showed near-significant alterations. However, unlike our cohort ($>$10 months post-injury), their acute-stage patients likely exhibited more pronounced abnormalities. Chronic PPCS lesions may be too subtle for TBSS to detect, whereas tract-based methods retain enough local sensitivity to capture lingering effects.

\noindent Taken together, our findings emphasize that the optimal diffusion analysis strategy depends not only on the acquisition and modeling choices but also on the underlying pathology and sample characteristics. TBSS and ROI-based approaches differ in their ability to detect diffuse versus focal white-matter changes, and the relative performance of DTI and DKI likewise varies according to the scale and nature of tissue injury. In this context, chronic PPCS represents a particularly challenging case for TBSS---characterized by small samples, focal and spatially variable lesions, and subtle microstructural changes that rarely align across subjects---whereas chronic SLE exemplifies conditions under which TBSS is most effective, with larger cohorts, diffuse white-matter involvement, and strong, spatially coherent effects. Our results emphasize that methodological choice should not be secondary to acquisition or modeling decisions but must be tailored to the pathology under investigation. Recognizing these dependencies is essential for designing adequately powered studies and for interpreting negative or heterogeneous findings in small clinical cohorts. Beyond methodological implications, such evidence-based selection of diffusion analysis frameworks could improve reproducibility across research centers and enhance the diagnostic and prognostic utility of diffusion MRI in disorders such as mild TBI, SLE, and other neuroinflammatory diseases.

\noindent Furthermore, our TBSS analysis of WM pathology in SLE showed that the choice of the diffusion protocol was more influential for the sensitivity than the magnetic field strength and the processing pipeline. For all the common diffusion parameters that we examined (FA, MD, AD and RD), the two DTI protocols (3T-DTI and 7T-DTI) yielded higher sensitivity results than the 3T-DKI one. Moreover, except for RK, the kurtosis parameters derived by 3T-DKI (MK and AK) did not provide as high a sensitivity as RD and FA, the two most sensitive tensor parameters. FA was the parameter displaying the highest sensitivity in all the protocols. Using 3T-DTI to compute FA, 44\% of all voxels on the skeleton exhibited significantly different values in SLE patients than in controls. The corresponding values for an FA derived by 3T-DKI and 7T-DTI were 19\% and 23\% respectively. A post-hoc evaluation, in which the healthy controls were split into two groups and TBSS was run between them to check for false positives, supported the reliability of the differences reported between patients and controls, since no significant white matter differences were observed between the two control groups. Sensitivity from 3T-DTI was consistently higher than with 7T-DTI. These results were consistent across the seven tested processing pipelines. Overall, the results on analysing the sensitivity of dMRI to SLE pathology using TBSS corroborate our initial findings on the same study using an ROI-based sensitivity analysis \citep{kornaropoulos2022sensitivity}, where no increase in sensitivity came along with the use of more demanding acquisitions (DKI and 7T).

\noindent Another dimension concerns the choice of diffusion model. In our data, DKI showed lower sensitivity than DTI to white-matter pathology in SLE, regardless of whether ROI- or TBSS-based analyses were used. This observation is consistent with reports in other conditions, including schizophrenia \citep{zhu2015performances}, mild TBI \citep{muftuler2020serial}, and major depression \citep{kamiya2018diffusional}. Interestingly, although DTI also yielded higher sensitivity than DKI in the PPCS cohort, this difference was less pronounced than in SLE. The Gaussian diffusion assumption underlying DTI may be sufficient to capture the widespread, homogeneous abnormalities observed in SLE, whereas PPCS involves non-Gaussian diffusion processes---such as axonal beading, shearing, and microstructural heterogeneity---that should, in principle, be better characterized by DKI metrics \citep{jensen2005diffusional,fieremans2011white}. In contrast, several studies have reported higher diagnostic sensitivity of DKI compared with DTI in pathologies with well-defined microstructural alterations, such as Parkinson's disease \citep{kamagata2014preliminary}, Alzheimer's disease \citep{chen2017automated}, and pediatric temporal lobe epilepsy \citep{gao2012diffusion}. Collectively, these findings indicate that the relative advantage of DKI over DTI may depend on the specific pattern and scale of white-matter injury. Beyond pathology-specific factors, methodological aspects could also play a role. In our dataset, the DKI sequence had lower spatial resolution than the DTI acquisition, and voxel size is known to influence group-level DTI results \citep{barrio2015impact}. Consistent with this, our smoothing experiments demonstrated that reducing spatial resolution generally diminished sensitivity. Whether DKI would retain lower sensitivity than DTI when matched for resolution and SNR remains an open question that warrants systematic investigation.

\noindent The field-strength analysis using either the ROI- or the TBSS-based sensitivity approach showed higher sensitivity for 3T-DTI than 7T-DTI. It could be argued that this was due to the shorter scan time of the 7T-DTI protocol, which featured fewer diffusion encoding directions than the 3T-DTI protocol (30 versus 64, respectively). Earlier studies, however, have shown that there is no substantial increase in the accuracy of diffusion parameter estimation above approximately 25 encoding directions \citep{correia2009looking}. Another factor that could have reduced the sensitivity of 7T-DTI compared to 3T-DTI is the shorter T2* relaxation time at 7T. A consequence of that is a reduction of the intensity of k-space lines far from the center and a subsequent image blurring (i.e.\ smoothing) during acquisition. Furthermore, 7T-DTI has a higher baseline SNR than 3T-DTI, which could allow the use of a higher spatial resolution that in turn may increase sensitivity \citep{de2019characterizing}.

\noindent With respect to the processing pipelines, we found that the incorporation of any processing step increased the sensitivity, but not substantially, with the exception of eddy current and motion correction in 3T-DKI. On top of this, smoothing reduced sensitivity by up to 12\%. The detrimental effect of smoothing had been observed in the previous ROI-based study too \citep{kornaropoulos2022sensitivity}.

\noindent The study also has a few limitations, of which we consider three here. First of all, the pipelines we considered included neither outlier detection \citep{ades2018evaluation} nor harmonization \citep{fortin2017harmonization}. Regarding the former, however, manual inspection did not reveal any clear outliers in our data, while harmonization was not necessary for our study, since we ran separate analyses on the different protocols and investigated between-group effect sizes per protocol: Cohen's $d$ is a metric of groupwise differences in the mean normalized by the pooled SD. When computed per protocol, inter-protocol biases do not need to be considered. Second, the resolution of 3T-DKI was lower than that of 3T-DTI (2.3\,mm isotropic vs.\ 2.0\,mm isotropic). As mentioned above, the poorer resolution might explain the lower effect sizes in the SLE study with DKI, in particular, given that smoothing led to reduced effect sizes. Third, not all subjects were scanned with all four protocols, which might have induced some systematic sample-related differences in the results. However, the majority of subjects in each study (SLE or PPCS) overlapped and the overall picture did not change when including only matching subjects (regarding the SLE study we refer the interested reader to the Supplementary Material of \citealp{kornaropoulos2022sensitivity}).

\section{Conclusion}

\noindent In conclusion, the choice of dMRI sensitivity analysis approach (i.e.\ whether one opts for a TBSS- or a ROI-based approach) may introduce bias into the outcomes, depending on the type of data analyzed. Moreover, the choice of the diffusion protocol had a greater impact on sensitivity to both SLE and PPCS WM pathology than that of the magnetic field strength or the post-processing strategy. The use of more demanding acquisitions (DKI and 7T) did not yield any substantial increase in sensitivity. Incorporation of eddy current and motion correction as well as correction for Gibbs artefacts increased sensitivity to SLE pathology. Denoising was of marginal utility. Smoothing prior to parameter estimation was detrimental, indicating that high resolution is important for dMRI. Taking this into account in protocol design may well improve the sensitivity of future DKI and 7T dMRI studies.

\section*{Conflict of Interest}
The authors declare that the research was conducted in the absence of any commercial or financial relationships that could be construed as a potential conflict of interest.

\section*{Author Contributions}
EK contributed to the development of the pipelines, the conception and design of the analysis, the interpretation of the results, analysis of data and drafted the work. TR contributed to the acquisition of the SLE data for the work and the revision of the work for important intellectual content. AG contributed to the acquisition of the SRC data for the work. LK contributed to the revision of the work for important intellectual content. NM supervised the acquisition of the SRC data and contributed to the interpretation of data for the work and the revision of the work for important intellectual content. PS supervised the acquisition of the SLE data for the work, the interpretation of data for the work and the revision of the work for important intellectual content. MN contributed to the conception and design of the analysis, the interpretation of data for the work and the revision of the work for important intellectual content.

\section*{Funding}
The study was supported by Regional Research Funds (RegSkane-824651) (PS), SUS Foundation and Donations funds (PS), Alfred Österlund Foundation (PS), Swedish Rheumatism Association R-568371 (PS), and King Gustaf V’s 80-year Foundation (FAI-2019-0559) (PS).

\section*{Data Availability Statement}
The data analyzed in this study are not publicly available due to ethical restrictions, but are available from the corresponding author upon reasonable request.

\bibliographystyle{apalike}
\bibliography{literature_db}

\clearpage

\begin{table}[p]
    \centering
    \caption{Demographics and image acquisition parameters. PAT: patients (SLE or PPCS); HC: healthy controls. Ages are given as mean (SD) in years; resolution is isotropic voxel size.}
    \label{tab:image_acquisition}
    \small
    \begin{tabularx}{\textwidth}{l l c c c >{\raggedright\arraybackslash}X c}
        \toprule
        \textbf{Protocol} & \textbf{Group} & \textbf{\#PAT, \#HC} & \textbf{Age PAT, HC} & \textbf{Res. (mm)} & \textbf{b-values in s/mm$^2$ (\# directions)} & \textbf{TR/TE (ms)} \\
        \midrule
        \textbf{\color{NavyBlue}3T-DTI}     & SLE  & 63, 20 & 36 (9), 37 (9)  & 2.0 & 0 (8), 1000 (64) & 7300/73 \\
        \textbf{\color{YellowOrange}3T-DKI} & SLE  & 56, 20 & 35 (9), 37 (9)  & 2.3 & 0 (3), 250 (6), 500 (6), 1000 (20), 2750 (30) & 7500/103 \\
        \textbf{\color{OliveGreen}7T-DTI}   & SLE  & 54, 21 & 40 (9), 40 (10) & 2.0 & 0 (3), 1000 (30) & 8816/62 \\
        \textbf{\color{OliveGreen}7T-DTI}   & PPCS & 20, 21 & 25 (4), 26 (4)  & 2.0 & 0 (3), 100 (6), 1000 (30) & 9200/65 \\
        \textbf{\color{BrickRed}7T-DKI}     & PPCS & 20, 21 & 25 (4), 26 (4)  & 2.0 & 0 (3), 100 (6), 500 (6), 1000 (10), 2000 (30) & 9800/76 \\
        \bottomrule
    \end{tabularx}
\end{table}

\begin{table}[p]
\centering
\caption{White matter tracts for which this study found significant differences (effect size $d>0.54$) between healthy controls and SLE patients. Tracts in bold exceeded an effect size of 0.84 (Bonferroni-corrected threshold).}
\label{tab:tracts_sle}
\small
\begin{tabularx}{\textwidth}{l l X}
\toprule
\textbf{Parameter} & \textbf{Protocol} & \textbf{WM tracts with effect size $>0.54$} \\
\midrule
FA & \textbf{\color{NavyBlue}3T-DTI} & \textbf{Genu, Anterior thalamic radiation, Striato-prefrontal, Thalamo-prefrontal, Superior longitudinal fascicle III, Rostral body, Arcuate fascicle}, Superior longitudinal fascicle I, Inferior occipito-frontal fascicle, Inferior longitudinal fascicle, Striato-fronto-orbital, Anterior midbody, Posterior midbody \\
   & \textbf{\color{YellowOrange}3T-DKI} & Posterior midbody \\
   & \textbf{\color{OliveGreen}7T-DTI} & Striato-premotor, Striato-prefrontal, Superior longitudinal fascicle II, Striato-fronto-orbital, Striato-precentral, Superior longitudinal fascicle III \\
\midrule
MD & \textbf{\color{NavyBlue}3T-DTI} & Posterior midbody \\
\midrule
RD & \textbf{\color{NavyBlue}3T-DTI} & Genu, Anterior thalamic radiation, Striato-prefrontal, Thalamo-prefrontal, Superior longitudinal fascicle III, Rostral body, Arcuate fascicle, Anterior midbody, Posterior midbody \\
   & \textbf{\color{YellowOrange}3T-DKI} & Anterior midbody, Posterior midbody \\
   & \textbf{\color{OliveGreen}7T-DTI} & Corticospinal tract \\
\bottomrule
\end{tabularx}
\end{table}

\begin{table}[p]
\centering
\caption{White matter tracts for which this study found significant differences (effect size $d>0.64$) between healthy controls and PPCS athletes. Tracts in bold exceeded an effect size of 0.84 (Bonferroni-corrected threshold).}
\label{tab:tracts_ppcs}
\small
\begin{tabularx}{\textwidth}{l l X}
\toprule
\textbf{Parameter} & \textbf{Protocol} & \textbf{WM tracts with effect size $>0.64$} \\
\midrule
FA & \textbf{\color{OliveGreen}7T-DTI} & Middle longitudinal fascicle \\
   & \textbf{\color{BrickRed}7T-DKI} & \textbf{Superior longitudinal fascicle II, Fronto-pontine tract, Striato-parietal, Thalamo-parietal, Parieto-occipital pontine, Corpus callosum, Middle longitudinal fascicle}, Genu, Cingulum \\
\midrule
MD & \textbf{\color{OliveGreen}7T-DTI} & \textbf{Middle longitudinal fascicle, Inferior occipito-frontal fascicle, Striato-parietal, Superior longitudinal fascicle II}, Corpus callosum, Cingulum, Superior longitudinal fascicle I, Striato-prefrontal, Genu \\
   & \textbf{\color{BrickRed}7T-DKI} & \textbf{Superior longitudinal fascicle II, Fronto-pontine tract, Striato-parietal, Thalamo-parietal, Parieto-occipital pontine, Corpus callosum, Middle longitudinal fascicle}, Cingulum \\
\midrule
AD & \textbf{\color{OliveGreen}7T-DTI} & Middle longitudinal fascicle, Striato-parietal, Corpus callosum \\
\midrule
RD & \textbf{\color{OliveGreen}7T-DTI} & \textbf{Middle longitudinal fascicle, Striato-parietal, Inferior occipito-frontal fascicle}, Corpus callosum, Cingulum, Superior longitudinal fascicle I, Striato-prefrontal, Genu \\
   & \textbf{\color{BrickRed}7T-DKI} & \textbf{Superior longitudinal fascicle II, Fronto-pontine tract, Striato-parietal, Thalamo-parietal, Parieto-occipital pontine, Corpus callosum, Middle longitudinal fascicle}, Cingulum \\
\midrule
AK & \textbf{\color{BrickRed}7T-DKI} & \textbf{Middle longitudinal fascicle, Corpus callosum, Parieto-occipital pontine, Cingulum}, Genu \\
\bottomrule
\end{tabularx}
\end{table}

\begin{table}[p]
    \centering
    \caption{Articles comparing TBSS- to ROI-based dMRI sensitivity analysis. SRC: sport-related concussion; PD: Parkinson's disease; BD: bipolar disorder.}
    \label{tab:articles_tbss_and_roi}
    \small
    \begin{tabularx}{\textwidth}{l l l X X}
        \toprule
        \textbf{Reference} & \textbf{Pathology} & \textbf{Preprocessing} & \textbf{Sensitivity metric, ROI-based} & \textbf{Sensitivity metric, TBSS} \\
        \midrule
        \citet{grinberg2017diffusion} & Aging    & Eddy          & Cohen's $d$     & Qualitative assessment of p-value maps (FSL \textit{randomise}) \\
        \citet{xiong2019brain}        & Diabetes & Eddy          & t-test (SPSS)   & Qualitative assessment of p-value maps (FSL \textit{randomise}) \\
        \citet{karlsen2019diffusion}  & SRC      & Eddy \& TOPUP & ANCOVA (SPSS)   & Qualitative assessment of p-value maps (FSL \textit{randomise}) \\
        \citet{sejnoha2020patterns}   & PD       & Eddy \& TOPUP & PALM (FSL)      & Qualitative assessment of p-value maps (FSL \textit{randomise}) \\
        \citet{yang2021white}         & BD       & PANDA         & Cohen's $d$     & GLM \\
        \bottomrule
    \end{tabularx}
\end{table}

\clearpage

\end{document}